\documentclass[twocolumn,showpacs,preprintnumbers,amsmath,amssymb]{revtex4}

\usepackage{graphicx}
\usepackage{dcolumn}
\usepackage{bm}
\usepackage{subfigure}
\usepackage{amsmath}
\usepackage{wasysym}

\newcommand{\hz}{\,{\rm Hz}}
\newcommand{\khz}{\,{\rm kHz}}
\newcommand{\Mhz}{\,{\rm MHz}}

\newcommand{\rb}{$^{87}$Rb\;}
\newcommand{\uw}{\,\mu{\rm W}}

\newcommand{\pw}{\,{\rm pW}}
\newcommand{\us}{\,\mu{\rm s}}
\newcommand{\ms}{\,{\rm ms}}
\newcommand{\s}{\,{\rm s}}
\newcommand{\nG}{\,{\rm nG}}
\newcommand{\pG}{\,{\rm pG}}
\newcommand{\uG}{\,\mu{\rm G}}
\newcommand{\rd}{{\rm d}}
\newcommand{\cm}{\,{\rm cm}}

\newcommand{\rp}{\right )}
\newcommand{\lp}{\left (}

\newcommand{\beq}{\begin{equation}}
\newcommand{\eeq}{\end{equation}}
\newcommand{\bea}{\begin{eqnarray}}
\newcommand{\eea}{\end{eqnarray}}
\newcommand{\inv}[1]{\frac{1}{#1}}
\newcommand{\comment}[1]{}

\begin{document}



\title{ Robust,  High-speed, All-Optical Atomic Magnetometer }

\author{J.\ M.\ Higbie}
\affiliation{%
Department of Physics, University of California, 
Berkeley, CA  94720
}
\author{E.\ Corsini}

\affiliation{%
Department of Physics, University of California, 
Berkeley, CA  94720
}%

\author{D.\ Budker}
\email{budker@berkeley.edu} \affiliation{Department of
Physics, University of California, Berkeley, CA 94720-7300}
\affiliation{Nuclear Science Division, Lawrence Berkeley National
Laboratory, Berkeley CA 94720}

\date{\today}

\begin{abstract}
A self-oscillating magnetometer based on the nonlinear magneto-optical
rotation effect with separate modulated pump and unmodulated probe beams
is demonstrated. This device possesses a bandwidth exceeding $1\khz$. 
Pump and probe are delivered by optical fiber, facilitating miniaturization
and modularization. The magnetometer has been operated both with vertical-cavity 
surface-emitting lasers (VCSELs), which are well suited to portable 
applications,
and with conventional edge-emitting diode lasers.
A sensitivity of around
$3\,{\rm nG}$ is achieved for a measurement time of $1\s$. 
\end{abstract}

\pacs{07.55.Ge,
33.55.Ad
}                             
\maketitle

\section{\label{sec:intro}Introduction}
Considerable progress has been made in recent years in atomic magnetometry,
including the achievement of subfemtotesla sensitivity \cite{Kominis2003}, 
the application of coherent dark states to magnetometry 
\cite{Nagel1998,Affolderbach2002}, development of sensitive atomic
magnetometers for biological applications \cite{Bison2003,Xia2006}
and for fundamental physics 
\cite{murthy1989,berglund1995,youdin1996,romalis2001,amini2003,Groeger2006},
the introduction of chip-scale atomic magnetometers \cite{schwindt2004},
and
the development of nonlinear magneto-optical rotation (NMOR) 
using modulated light \cite{Budker2000}
with its subsequent 
demonstration in the geomagnetic field range \cite{Ledbetter2006}.
An important potential application is   
to space-based measurements, including measurements of planetary fields,
of the solar-wind current, and of magnetic fields in deep space \cite{acuna2002}.
As these sensitive magnetometric techniques move from the laboratory 
to applications in the field and in space, significant new demands
will naturally be placed on their robustness, size, weight,
and power consumption. 
An attractive approach that
addresses many of these demands
is the self-oscillating magnetometer configuration, originally proposed by Bloom 
\cite{Bloom1962}. 
In this configuration, the detected optical signal 
\comment{NEED EARLIER DEFINITION OF TERMS?} at the Larmor
frequency or a harmonic, is used to drive the magnetic resonance, either with RF coils or, 
as in the present work, by optical excitation of the resonance.
When this resonant drive is amplified sufficiently, the oscillation builds up 
spontaneously (seeded by
noise) at the resonant frequency. 
The spontaneous nature of the oscillation obviates the necessity of
 slow preliminary scans, which would otherwise be required
 to search for the resonance and lock to 
 the desired line. Moreover, the size, weight, and power
consumption of a self-oscillating device all benefit from the simplification
of the electronics that results from not requiring a local oscillator or
lock-in amplifier.

\begin{figure}
\mbox{
\includegraphics[scale=1]{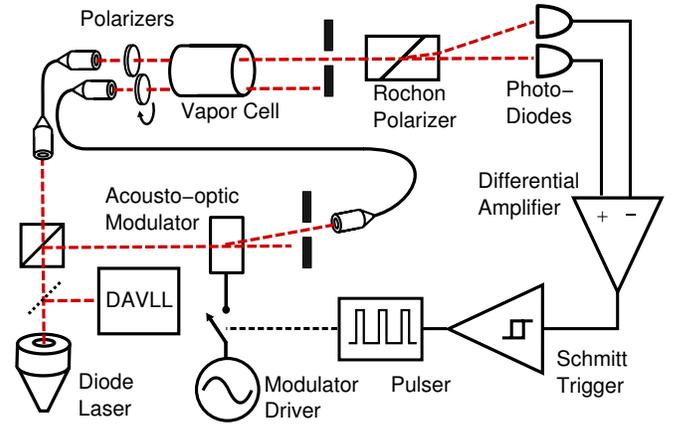}
}
\caption{\label{fig_diagram} Diagram of experimental setup.
The output of a diode laser is split into pump and probe beams.
The laser is frequency-stabilized approximately $300\Mhz$ below the 
center of the \rb D2 Doppler-broadened line center via a dichroic atomic vapor
laser lock (DAVLL). The pump beam is amplitude-modulated by
an acousto-optic modulator (AOM). Both pump and probe are delivered
by optical fiber to a paraffin-coated vapor cell. Separate polarizers allow
independent control of pump and probe polarization. The polarization
rotation of the probe beam is analyzed in a balanced polarimeter
consisting of a Rochon polarizer followed by a
pair of photodiodes and an amplifier. A zero-crossing
detector (Schmitt trigger) and pulser control the
pump power via the AOM, closing the loop and sustaining self-oscillation.
In an alternate configuration, separate VCSELs on the \rb D1 F=2
line supplied light for pump
and probe.
}
\end{figure}

A self-oscillating magnetometer
based on NMOR 
was recently reported by Schwindt 
et al. \cite{schwindt2005},
in whose work 
 the functions of pumping and probing were fulfilled
by a single frequency-modulated laser beam. 
As a result, the detected signal was a product both of the rotating atomic alignment
and of the modulated detuning, resulting in a complicated waveform 
that required significant electronic processing
before being suitable for feeding back to the laser modulation, 
as required in the self-oscillating scheme.

In the present paper, we present a simple alternative arrangement
that avoids many of the difficulties encountered in the single-beam experiment. Indeed,
 by the use of two laser
beams--a modulated pump and an unmodulated probe--the optical-rotation 
signal may be made accurately sinusoidal, avoiding
the complexity of digital or other variable-frequency filters in the feedback loop. 
Importantly, the use of two beams
also permits optical adjustment of the relative phase of the 
detected signal and the driving modulation by changing
the angle between their respective linear polarizations. For 
magnetometry at large bias field and requiring a wide range of fields, 
this optical tuning of the feedback-loop phase promises both 
good long-term stability and much greater 
uniformity with respect to frequency
 than can readily be obtained with an electronic phase shift.

\section{NMOR Resonance}

Detailed discussions of zero-field NMOR resonances \cite{Budker1998,Budker2000},
as well as of the additional
finite-field resonances that occur when the pumping laser light  
is frequency-modulated (FM)
\cite{Budker2002}
or amplitude-modulated (AM) \cite{Gawlik2006}, have been presented 
in prior work. An NMOR resonance occurs when
optical pumping causes an atomic vapor 
to become dichroic (or 
birefringent), so that subsequent probe light experiences polarization 
rotation. 
For the resonances considered
in this work, both pump and probe are linearly polarized and therefore 
primarily produce and detect atomic alignment ($\Delta m = 2$ coherences). The 
magnetic-field dependence originates from the fact that the atomic 
spins undergo Larmor precession, 
so that weak optical pumping can only produce a macroscopic 
alignment when the Larmor precession frequency is small
compared to the spin relaxation rate, or alternatively when 
pumping is nearly synchronous with precession, as in FM or AM NMOR. 

If the optical-pumping rate  is modulated at a frequency $\nu$, then
the optical-rotation angle of the probe polarization will in general
also oscillate at frequency $\nu$. 
If this frequency is scanned
across the resonance (in open-loop configuration, i.e. with no feedback from
the optical-rotation signal),
then the NMOR resonance will manifest itself
as a resonant peak in the rotation-angle amplitude of the
probe polarization on the output.
 Assuming the in-going probe and pump polarizations 
to be parallel,  the  amplitude
and phase of the observed rotation signal  can be described by the complex Lorentzian
$
   (\delta - i \gamma /2)^{-1}$,
where $\delta \equiv 2\pi (\nu-2\nu_{\rm L})$ is the detuning from resonance, $\gamma$
is the full width (in modulation frequency) at half maximum of the resonance, and 
$\nu_{\rm L}$ is the Larmor frequency. The phase shift 
relative to the pump modulation as a function of $\delta$ is seen
by taking the argument of this complex Lorentzian to be
\beq
\label{phase_freq_rel}
\phi = \frac{\pi}{2}+ \tan^{-1}\lp \frac{2\delta}{\gamma} \rp
.\eeq 
This elementary relation, which is the same as for a damped harmonic oscillator,
will be referred to frequently in subsequent sections.

\section{\label{sec:apparat}Apparatus}
The experimental apparatus, shown schematically in Fig.\
\ref{fig_diagram}, consists of a cylindrical paraffin-coated \rb vapor cell $2\cm$
in diameter and length
traversed by linearly-polarized pump and probe laser beams. 
These beams were supplied  
by
a single external-cavity diode laser on the D2 line of rubidium, 
frequency-stabilized
$\sim 300\Mhz$ below the center of the $F=2\longrightarrow F'$ Doppler-broadened
 line by means of a dichroic atomic
vapor laser lock \cite{Corwin1998,Yashchuk2000}. 
The probe beam was left unmodulated, while the pump was amplitude modulated 
with an acousto-optic modulator (AOM). 
Pump and probe were delivered to the cell by separate polarization-maintaining
fibers.
After exiting the cell, the pump beam was blocked and the probe analyzed
by a balanced polarimeter consisting of a Rochon polarizing beam-splitter 
and a pair of photodiodes.
The difference photocurrent was amplified with a low-noise transimpedance amplifier
(Stanford Research Model SR570) and passed through a resonant LC filter
centered at $20\khz$ with a bandwidth of $11\khz$, much wider than either the NMOR 
resonance ($\sim 80\hz$) or the desired magnetometer bandwidth ($\sim 1\khz$).
This filter reduced jitter in the frequency-counter readings, but is not necessary
in principle.
The pump modulation was derived from this amplified signal, closing the feedback loop,
by triggering a pulse generator on the negative-going 
zero-crossings of the signal, and allowing these
pulses to switch on and off the radiofrequency power delivered to the AOM.
The pulse duty cycle was approximately $15\%$.
For characterization of the magnetometer in the laboratory, the vapor cell was
placed in a three-layer cylindrical magnetic shield, provided with internal coils
for the generation of a stable, well-defined magnetic bias field and gradients. The \rb density 
in the cell was maintained at an elevated value ($\approx 5\times 10^{10}\cm^{-3}$ as
measured by absorption) by heating the 
interior of the magnetic shields to around $40^\circ$C
with a forced-air heat exchanger. 

The photodiode signal was monitored with an oscilloscope and a frequency counter (Stanford Research
Model SR620). 
Provided the trigger threshold of the pulse generator was close enough to zero (i.e. within a few
times the noise level of the signal), oscillation would occur spontaneously when the 
loop was closed
at a frequency set by the magnetic field.
Optimum settings for the magnetometer sensitivity were found to be 
approximately $7\uw$ mean incident pump power, $7\uw$ continuous incident probe power,
 and optical absorption
of around $60\%$ at the lock point. 
A sensitivity of $3\nG$ was achieved for a measurement time 
of $1\s$ at these settings, as discussed in detail in section \ref{performance_sect}.

Several alternative configurations were also implemented. In place of the balanced polarimeter
described above, a configuration consisting of a
polarizer nearly orthogonal to the unrotated probe polarization
 followed by a large-area avalanche photodiode (APD)
module was employed. This configuration has high detection bandwidth, but suffers
from lower common-mode noise rejection and greater sensitivity to stray light. Moreover, the
excess noise factor of the APD module prevents attaining shot-noise-limited operation.
With the APD module, self-oscillation at frequencies up to $\sim 1\Mhz$ was achieved.
In another configuration, frequency modulation of the pump laser was employed. This configuration
worked well, but the laser frequency lock point was found to depend subtly on the 
state of self-oscillation of the magnetometer. For this reason, it was found 
preferable to employ amplitude modulation of the pump
laser
via an external modulator following the light pick off for laser frequency stabilization.
The magnetometer has moreover been operated
with two separate vertical-cavity surface-emitting diode lasers (VCSELs) as pump 
and probe on the D1 line of rubidium. The low power requirements, small size, and reliable
tuning of VCSELs render them appealing for use in miniaturized and portable magnetometers.
Amplitude modulation has also been performed with an inline fiber-optic Mach-Zehnder
interferometric modulator, which permits further miniaturization and considerably reduced
power consumption relative to the acousto-optic modulator.

\begin{figure}[h]
\includegraphics[scale=.6]{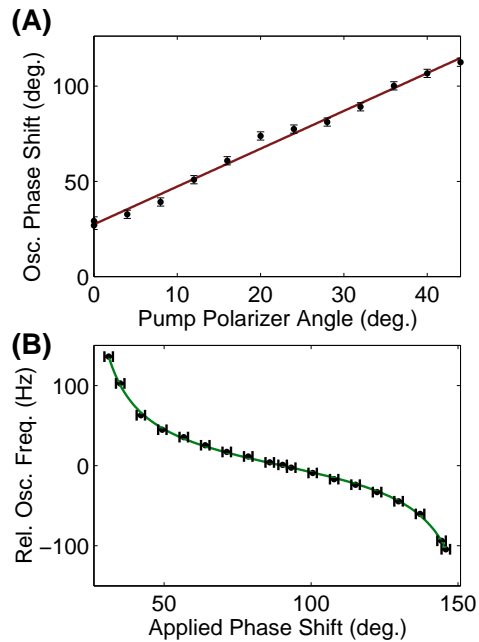}
\caption{\label{fig_phase} 
Demonstration of all-optical phase shift. The pump polarization 
was rotated while the probe polarization was held fixed; for each angular
setting a pulse delay (equivalent to an electronic phase shift) was applied to maintain the 
oscillator on resonance. In (a), this electronic phase shift is
plotted against the polarizer angle. The straight-line fit
gives a slope of $ 1.98 \pm 0.06$, in good agreement with the 
expected value of $2$ (see text).
In (b), the oscillator frequency as a function of phase shift 
is plotted. The curve agrees well with the expected form 
of Eq.\ \eqref{phase_freq_rel}, although the vertical asymptotes
of the tangent fit (solid line) do not occur at $0^\circ$ and $180^\circ$,
probably resulting from a small ($\sim 0.1\%$) residual 
contribution of scattered pump light 
to the optical rotation signal. 
The solid line is a fit to a tangent function. 
For phase shifts approaching $0^\circ$ or $180^\circ$, the gain becomes insufficient
to sustain self-oscillation.}
\end{figure}

\section{Optical Phase Shift}

To emphasize the advantages of the two-beam arrangement for the 
optical adjustment of the phase shift, we note that
optical pumping by linearly polarized light favors the preparation of
a state whose alignment symmetry axis 
is parallel to the pump polarization. At the center of the
NMOR resonance, therefore, where there is no phase lag between pumping and precession, the 
precessing axis of atomic alignment is parallel to the pump polarization at the moment
of maximal optical pumping in the limit of short pump pulses. 
Consequently, if the probe polarization is parallel to the
pump polarization, the probe polarization rotation signal will pass through zero at the same moment,
so that this optical rotation signal is $90^\circ$ out of phase with the pump modulation,
as seen in Eq. \eqref{phase_freq_rel}.
Thus the rotation signal must be shifted by $-90^\circ$ before being fed back as pump modulation
in order for coherent buildup at the central resonant frequency to occur. Deviations from this
phase shift will result in oscillation away from line center, 
in such a way as to maintain zero total phase
 around the loop, so long as the magnitude of the gain is sufficient to sustain oscillation at 
the shifted frequency.
As a result, precise control over this phase shift as a 
function of frequency is required to avoid (or compensate
for) systematic deviations of the oscillation frequency from $2\omega_L$ as a 
function of magnetic field.
Analog filter networks capable of generating accurate and stable $90^\circ$ 
phase shifts over a broad range of 
frequencies are difficult to construct. Digital phase shifters, 
although feasible, add complexity 
and power consumption and risk degradation of 
performance. 

The use of separate pump and probe
beams offers a natural and all-optical means of shifting the relative phase 
between modulation and optical rotation.
Indeed, since the rotation of the probe polarization is determined by 
the angle of the incident polarization
with respect to the axis of atomic alignment, which itself 
rotates uniformly in time, a fixed rotation of the 
incoming probe polarization is equivalent to a translation in time of the output signal,
 i.e., a phase shift. Since this phase shift is purely geometrical, it has no 
frequency dependence,
 and possesses the long-term stability of the mechanical mounting of the polarizers.

To demonstrate the optical phase shift of the polarization-rotation
signal, a measurement of the phase shift of the signal as a function of the pump polarizer angle 
was performed in the open-loop configuration, i.e., with an external frequency
source modulating the pump power
and with no feedback of the optical-rotation signal. The resulting curve reveals the 
expected linear dependence, 
as shown in Fig.\ \ref{fig_phase}(a). 
The observed slope is consistent with the value of $2$ expected from the fact that the 
optical-rotation signal undergoes two cycles as the atomic alignment rotates by $360^\circ$.
The effects of a phase shift in the feedback network
on the oscillation frequency of
the self-oscillating, closed-loop magnetometer are shown in Fig.\ \ref{fig_phase}(b). As expected,
the magnetometer adjusts its oscillation frequency so that the phase shift of the NMOR resonance
cancels that of the feedback network. Since the phase shift of the NMOR resonance is an arctangent
function of the detuning from resonance (similar to a damped harmonic oscillator),
this results in a change in 
oscillation frequency which is proportional to the tangent of the applied
phase shift, as seen in Fig.\ \ref{fig_phase}(b).

\begin{figure}[tbp]
\centering 
	\mbox{
  \includegraphics[scale=.5]{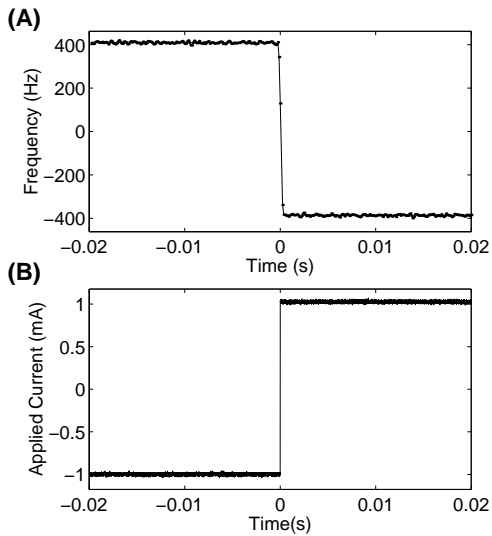}}
  \caption{\label{fieldstep2}Response of the magnetometer to a  $570\uG$
step
in magnetic field, in the presence of an over-all bias field of $\sim 14\,$mG.
This step is thus large compared to the resonance line width, but small compared
to the bias field.
  The self-oscillation waveform was recorded on a digital storage oscilloscope
  and subsequently fit to a sinusoid in overlapping time windows $500\us$ long, spaced
  by $125\us$. The resulting frequency is shown in part (a). In part (b), an oscilloscope
trace of the bias current  proportional to the field step is shown.
 }
\end{figure}

\begin{figure}[tbp]
\centering 
  \includegraphics[scale=.4]{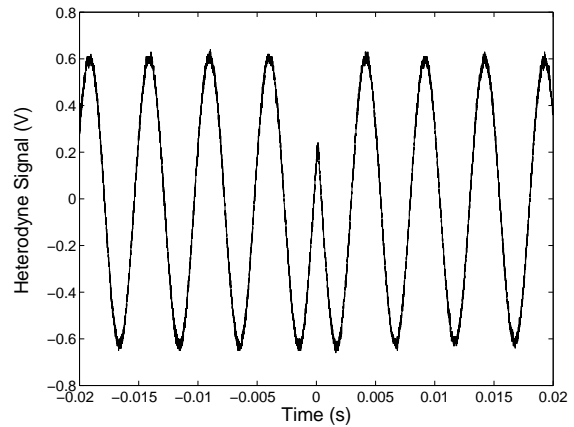}
  \caption{\label{fig_step_het} Heterodyne measurement 
  of magnetometer response to a field step. Shown here is the self-oscillating 
rotation signal mixed down to low frequency by 
a lock-in amplifier with its reference set to approximately twice the Larmor frequency.
 This heterodyned signal is shown for a period of approximately 
$40\ms$ during which a sudden change of $\approx 250\uG$
is made to the bias magnetic field.
The local oscillator is set midway between the two oscillation frequencies, so that
the heterodyne signal appears to reverse itself in time.}
\end{figure}

\section{Sinusoidal Output Signal}
In most past NMOR experiments, modulation was
 applied from a stable reference oscillator, and the 
rotation signal was demodulated by a lock-in amplifier at this reference 
frequency (or one of its harmonics).  On the basis of the
lock-in output, the reference frequency could be updated at a rate 
governed by the lock-in time constant, in order
to maintain the resonance condition in a slowly changing magnetic 
field. By contrast, in the self-oscillating scheme,
the measured rotation signal is fed back to the modulation input in 
place of the reference frequency, as shown in Fig.\ \ref{fig_diagram}. In order
 to reproduce self-consistently the effects of an external modulation, 
this signal must in general be phase shifted and
filtered so that it closely resembles the original reference frequency, 
as in the work of Ref. \cite{schwindt2005}.

Indeed, if the probe is frequency- or amplitude-modulated, as
is the case for a single-beam experiment,
then the observed rotation signal will be the product
of the rotation signal that would be observed with an 
unmodulated probe and a function which describes the modulation.
In the case of frequency modulation, for example, the 
observed rotation signal for an isolated line, Doppler broadened to 
a width $\Delta \nu_{\rm Doppler}$, would be approximately
\[
\phi_{FM}(t)\approx\phi_{un.}(t) e^{{-(\delta_0 - A 
\cos 2\pi\nu_{\rm mod} t)^2/\Delta \nu_{\rm Doppler}^2}},
\]
where $\phi_{un.}$ is the rotation that would be observed by an 
unmodulated probe passing through the same sample, $\delta_0$ is the 
mean detuning of the laser from resonance, $A$ the modulation 
amplitude, and $\nu_{\rm mod}$ the modulation frequency.
Similarly, in the case of pulsed amplitude modulation, 
the multiplicative modulation function would take
the form of a pulse train at the
modulation frequency. Since the atomic alignment 
is described by a rank 2 spherical tensor, and the
corresponding spin probability distribution is two-fold
symmetric (see, for example, Ref.\ \cite{Rochester2001}), the unmodulated
rotation signal is to good approximation sinusoidal at twice the 
Larmor frequency, $2\omega_L$. (Note that this argument
neglects the effects of alignment-to-orientation conversion
\cite{Bucker_AOC_2000}, which becomes important at relatively large
light powers.)
The over-all rotation signal detected by a modulated probe, however,
is in general highly non-sinusoidal. 
For stable and reproducible operation, such a signal would
almost certainly require filtering, which generically introduces
undesirable phase shifts. In contrast, the use of an 
unmodulated probe avoids this complication altogether. The 
detected rotation signal
is a near-perfect sinusoid (measurements indicate that the
higher harmonics are down by
more than $50\,$dB). Such a signal requires only amplification
to make it mimic the reference oscillator.

\section{High-speed Response}

In order to assess the bandwidth of the magnetometer,
the response to rapid changes in magnetic field was investigated
by applying a small modulation to the bias magnetic field. In one measurement,
a slow square-wave modulation was superimposed on the 
bias field via a separate Helmholtz coil inside
the magnetic shield. The self-oscillation signal was then recorded on an oscilloscope
and fit in each $500\us$ window to a sinusoid, with the results shown in Fig. \ref{fieldstep2}.
Tracking of the field step is quasi-instantaneous, without apparent overshoot or ringing.
The magnetometer response was also monitored by heterodyning the 
oscillation frequency 
with a fixed reference frequency on a lock-in amplifier, with the lock-in time constant set to
approximately the oscillation period ($\approx 50\us$) to remove the sum-frequency component.
The resulting low-frequency beat signal, which displayed large fractional frequency 
modulation, was also digitized and recorded on an oscilloscope. Inspection
of the waveforms so obtained revealed
the same sudden shift in the oscillation frequency as the 
magnetic field toggled between values (see Fig.\ \ref{fig_step_het}).
In a related experiment, the bias field received a small sinusoidal modulation, and
the power spectrum of the self-oscillation waveform was observed on a spectrum analyzer.
The sidebands were observed, offset from the oscillation (carrier) frequency 
by an amount equal to this bias-modulation frequency;
their relative power was equal to that expected if the 
oscillator
tracked the changing magnetic field with no delay or diminution of amplitude out to a
bias-modulation frequency of at least $1\khz$.

\begin{figure}
\includegraphics[scale=.4]{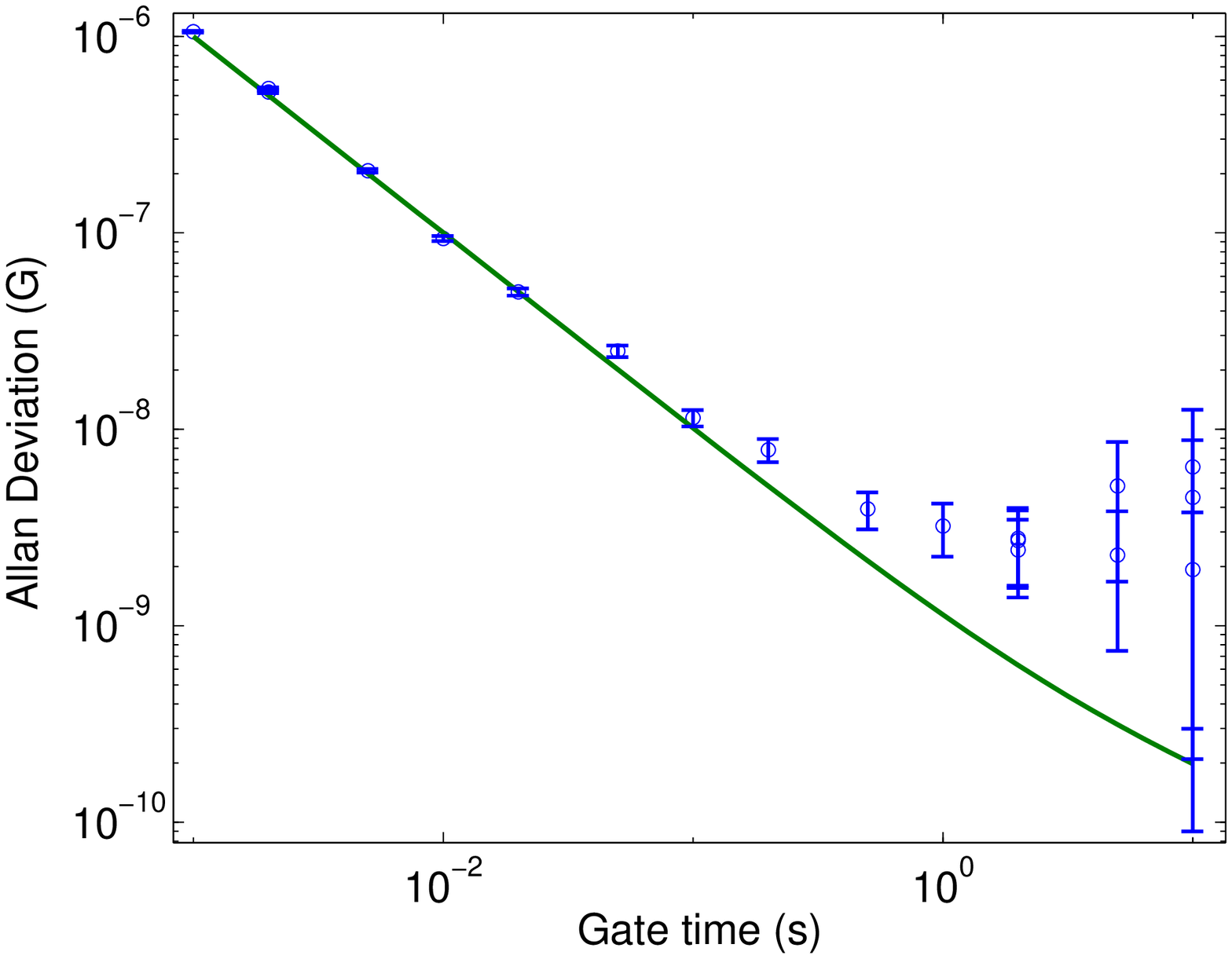}
\caption{\label{fig_allan} Allan deviation of magnetometer as obtained
from counter. \comment{Phase was
extracted from the heterodyned oscillation signal. In each measurement time $T$,
a frequency $\nu_i$ was computed as the least-squares slope $\frac{1}{2\pi}\frac{\rd \phi}{\rd t}$,
and the Allan deviation $\sigma_A(T) = \sqrt{\frac{1}{2} \langle (\nu_{i+1}-\nu_{i})^2\rangle}$
was computed} The solid line indicates the calculated sensitivity given by Eq.\
\eqref{eq_nu_unc_num}, showing good agreement for short measurement times.}
\end{figure}




\section{Comparison with Calculated Sensitivity}
\label{performance_sect}

To evaluate the performance of the magnetometer, it is useful
to calculate the performance that is expected from 
measurable system parameters
as a function of the measurement time.
The self-oscillating magnetometer is similar
in many respects to a maser, as was first pointed out in 
Ref.\ \cite{Bloom1962}; a treatment of noise in the
hydrogen maser system is given in Ref.\
\cite{Vanier}.
The read-out device envisioned here is a frequency counter that measures to high precision
the elapsed time between two zero-crossings of the rotation signal and reports
a frequency which is the integer number of zero-crossings divided by this time.

Noise in the magnetometer readings comes from several sources. These include
fundamental noise, such as atomic and photon shot noise, as well
as technical noise from electronics. Fluctuations in the measured field 
also appear as noise, but do not represent a failing of the magnetometer.
For a transmitted probe beam power of $P = 2.5\uw$, the optical shot noise 
for an ideal polarimeter, given in terms of the photon energy $E_{\rm ph}$ by 
$\sqrt{ 2 E_{\rm ph} P}$,
is $1.1\pw/\sqrt{\hz}$ or $0.55\,{\rm pA}/\sqrt{\hz}$
in terms of the differential photocurrent noise. Atomic shot noise is 
expected to contribute a comparable noise level 
for an optimized magnetometer \cite{Auzinsh2004}.
Observed amplifier noise 
is somewhat larger than the photon shot-noise
 level, or
$ 0.8\,{\rm pA}/\sqrt{\hz}$, and can be considered as white noise
over the relevant bandwidth ($\apprle 5\khz$) around the operation frequency
of $20\khz$. For comparison, the optimized self-oscillation signal amplitude,
hereafter denoted $I_0$, is $14\,\rm{nA}$.

Noise contributes to the uncertainty in the measurement of the self-oscillation frequency
 in two 
essential ways. First, it imparts random shifts to the times of the signal zero-crossings, 
resulting in jitter of the frequency-counter trigger times and of the corresponding reported
frequency. 
Second, noise within the NMOR resonance bandwidth drives the atomic resonance,
resulting in random drifts of 
the phase of oscillation and limiting the precision of the frequency measurement.
We will consider each of these in turn, both for the observed noise level and for
the photon-shot-noise limit.

Jitter of the counter triggers can be derived from a photocurrent difference
signal 
\[
I(t) = I_0 \sin(2\pi\nu_0 t) + \Delta I(t).\]
Here, $\Delta I(t)$ is additive noise, e.g.,  photon shot noise or  amplifier
noise, referred to the amplifier
input, i.e., expressed as a photocurrent, and $\nu_0$ is the mean self-oscillation frequency. 
 Provided that $\Delta I(t)$ is small, the zero-crossings of this signal
experience a r.m.s. fluctuation
\beq
\label{eq_tzero}
\Delta t_{\rm zero}= \frac{ \Delta I_{\rm rms}}{2\pi\nu_0 I_0}
.\eeq
If the open-loop noise spectrum is approximately white in the vicinity of $\nu_0$, then
the r.m.s. current is simply given by $\Delta I_{\rm rms}= \sqrt{2S_I  \nu_{\rm max}}$, where
$2\nu_{\rm max}$ is the bandwidth around $\nu_0$ defined, for instance, by a filter in the loop,
and $S_I$ is the single-sided current power spectral density, with units of ${\rm A}_{\rm rms}^2/\hz$. 
The r.m.s. deviation of the interval between two zero crossings $\Delta t_{\rm meas}$
is $\sqrt{2}$ times the 
deviation of each separately, so that the total uncertainty in the frequency
reading of the magnetometer due to trigger jitter is
\bea
\label{jitter_unc}
\Delta \nu_{\rm jitt} &=& \nu_0 {\Delta t_{\rm meas}}/{ T} \nonumber \\
&=&   \frac{(S_I  \nu_{\rm max})^{1/2}}{\pi T I_0}
,\eea
where $T$ is the measurement duration.

The second effect of noise on the operation of the magnetometer is due
to the feedback network that produces self-oscillation.
As shown below, noise that is within the linewidth of the NMOR
resonance mimics a fluctuating phase shift in the self-oscillating loop.
This phase shift produces random fluctuations in the oscillation frequency,
inducing diffusion
of the over-all phase of oscillation.
For notational simplicity, let us first consider a single frequency component of the noise
at a frequency $\nu_0+\nu_{\rm off}$, where $\nu_{\rm off}$ is referred to as the offset frequency.
For an open-loop magnetometer, i.e.,  with modulation supplied by an external frequency source
tuned to the NMOR resonance and
no feedback of the optical rotation signal,
the resulting photocurrent difference signal would be
\bea
\label{open_loop_sig}
I(t)&=& I_0 \sin (2\pi \nu_0 t) + \epsilon \sin (2\pi (\nu_0 + \nu_{\rm off}) t )
\nonumber\\
&\approx & I_0 \sin \lp 2\pi \nu_0 t + \frac{\epsilon}{I_0}\sin (2\pi  \nu_{\rm off} t)
\rp
,\eea
where $\epsilon$ is taken to be small, and a term contributing only to amplitude modulation
of the signal has been neglected. The phase of the noise component has been chosen arbitrarily
but plays no role in what follows.
Equation \eqref{open_loop_sig} shows that the effect of this noise component is to modulate
the phase of the open-loop signal with an amplitude $\epsilon/I_0$ at frequency $\nu_{\rm off}$.
When the self-oscillating loop is closed, the oscillation  responds to this noise-induced
phase shift by shifting the frequency of oscillation in such a way as to keep the net phase shift
around the loop zero. The magnitude of the resulting frequency modulation of the self-oscillating
signal is readily calculated from Eq.\ \eqref{phase_freq_rel},
which may be approximated over the central portion of the resonance
as linear, i.e., $\nu_{\rm off} \approx \gamma \phi/4\pi$, so that the amplitude of
the induced frequency modulation is
$
\gamma \epsilon/ 4\pi I_0
$. 
The resulting self-oscillation signal is
\bea
I(t)&=& I_0 \sin \int_{0}^t \rd t' \lp2\pi \nu_0 + 
\frac{\gamma \epsilon}{ 2 I_0 }\sin 2\pi \nu_{\rm off}t' \rp
\nonumber \\
&=&I_0 \sin \lp 2\pi \nu_0 t - 
\frac{\gamma \epsilon}{ 4\pi I_0 \nu_{\rm off} }\cos 2\pi \nu_{\rm off}t \rp
,\eea
where the global phase of oscillation has been taken to be zero at $t=0$.
In analogy to Eq.\ \eqref{eq_tzero}, the r.m.s. deviation of a zero-crossing 
of this signal at a randomly chosen time
is 
\[
\Delta t_{\rm zero}=
\frac{\gamma \epsilon}{ 4\pi I_0 \nu_{\rm off}} \inv{2\pi \nu_0 \sqrt{2}}
,\]
and the corresponding frequency uncertainty from phase diffusion
is
\beq
\label{deltanu_single}
\Delta \nu_{\rm diff}
=
\frac{\gamma \epsilon}{ 4\pi I_0 \nu_{\rm off}} \inv{2\pi T }
.\eeq
Although this expression diverges for $\nu_{\rm off}\rightarrow 0$, in reality a
measurement lasting a time $T$ imposes an effective low-frequency cut-off of $\approx 1/T$.
To take into account the fact that the noise contains many incoherent spectral 
components, rather than a single monochromatic component, one must add these components in
quadrature. Thus, it is sufficient to replace
the original mean square current modulation $\epsilon^2/2$ at frequency $\nu_{\rm off}$, 
by the photocurrent
noise power $S_I \rd \nu_{\rm off}$ in a range $\rd \nu_{\rm off}$ around
frequency $\nu_0+\nu_{\rm off}$, 
and integrate the result over the appropriate range of frequency.
For a measurement time $T$, the minimum resolvable frequency is $\approx 1/T$. The
maximum frequency at which this noise-induced frequency shift occurs is $\approx \gamma/2$,
but for times $T \apprge 2/\gamma$, which in practice include all times for which
this type of noise is dominant, we may reasonably approximate the range of integration
as extending to infinity. Note that noise on either side of $\nu_0$ contributes equally, so that the
total integral is twice the integral evaluated on the positive side only.
Thus Eq.\ \eqref{deltanu_single} must be modified to
\bea
\label{diff_unc_new}
\Delta \nu_{\rm diff}
&\approx&
\frac{\gamma \sqrt{2}}{ 8\pi^2 I_0 T} 
\left | 2\int_{1/T}^{\infty} \rd \nu_{\rm off} \frac{S_I(\nu_0+\nu_{\rm off})
}{\nu_{\rm off}^2}
\right |^{1/2}
\nonumber \\
&\approx&
\frac{\gamma }{ 4\pi^2 \sqrt{T}} 
  \frac{S_I^{1/2}}{I_0}
.
\eea
The photocurrent noise spectrum $S_I$ has been assumed white over the range of integration
in evaluating the integral.

This intra-resonance noise can also be understood 
in terms of phase diffusion \cite{Lax1967}.
Indeed, although the frequency of the oscillation
is stabilized to twice the Larmor frequency in the
self-oscillating scheme, the global phase of oscillation
experiences no feedback or restoring force, and is thus 
subject to an effective Brownian motion or diffusion,
where shot noise or amplifier noise provides a
 stochastic driving force.
In this picture, the phase undergoes a random walk,
with a step size given by the r.m.s. noise within the resonance
and a time step given by $\sim 1/\gamma$.
Explicitly, the total phase noise within the resonance yields
a step size  
$\Delta \phi_{\rm step} \approx (S_I^{1/2}/I_0) \sqrt{\gamma/2\pi}$.
After a time $T$ the number of steps is 
approximately $N_{\rm steps}= \gamma T $,
so that the total frequency uncertainty given by 
\bea
\Delta \nu_{\rm diff} &\approx&  \frac{\Delta \phi_{\rm step}}{2\pi T} \sqrt{N_{\rm steps}}
\nonumber\\
&\approx& \inv{(2\pi)^{3/2}  }
\frac{S_I^{1/2}}{I_0} \frac{\gamma}{T^{1/2}}  
,\eea
which agrees up to a numerical factor with Eq.\ \eqref{diff_unc_new}

Numerically, for the experimental parameters
discussed above, we have $S_I^{1/2} / I_0 = 6\times 10^{-5}\,\hz^{-1/2}$,
$\gamma=2\pi\times 80\hz$,
and $\nu_{\rm max}=5\khz$ (the measured bandwidth of an LC filter
in the loop), so that
Eqs.  \eqref{jitter_unc} and \eqref{diff_unc_new} 
 imply a magnetic-field sensitivity
limit of
\beq
\label{eq_nu_unc_num}
\Delta B = 
\left \{ \lp \frac{1.0 \nG}{T/1\s}\rp^2  
+ 
\lp \frac{0.54 \nG}{ \sqrt{{T/1\s}}} \rp^2 
\right \}^{1/2}.
\eeq
Equation \eqref{eq_nu_unc_num} is in good agreement with the measured data for times
below $\sim 0.1\s$, as shown in Fig.\ \ref{fig_allan}, while above $\sim 0.1\s$ 
measurements fall short of calculated performance, as discussed below.

It should be noted that the initial $1/T$ dependence of the sensitivity
is non-optimal. If information about the phase during the entire duration $T$ were
retained and the frequency extracted (e.g. by least-squares fitting), then this
dependence could be improved by an additional factor of 
approximately $\sqrt{T_{\rm samp}/T}$, where $T_{\rm samp}$ 
is a sampling time, provided $1/T_{\rm samp}$ is smaller than the noise
bandwidth, e.g., the filter frequency
$\nu_{\rm max}$. The improved scaling would then be $\propto T^{-3/2}$. 
For measurement times $\apprge 1\s$, however,
this is not a limiting factor sufficient to outweigh the
frequency counter's considerable convenience.
The result of Eq.\ \eqref{eq_nu_unc_num} is within a factor of two
of the optical-shot-noise limit for the same system parameters, although
with a quieter photodiode amplifier, further optimization of light powers, detunings, and
\rb density should permit operation with narrower resonance lines
and considerable consequent improvement.

\comment{
It is also of interest to calculate the performance expected when the system
attains its fundamental noise limitation, i.e., optical shot noise, as should be possible
by use of a quieter amplifier.
If dominated by optical shot noise, the photocurrent noise
would drop to $\approx 0.55\,{\rm pA_{\rm rms}}$ for $2.5\uw$ transmitted probe
power, and the expected field sensitivity would scale accordingly,
yielding
\beq
\label{eq_shotnoise_unc}
\Delta B_{\rm shot} = \left \{ \lp \frac{0.69 \nG}{T/1\s}\rp^2  
+ 
\lp \frac{0.37 \nG}{ \sqrt{{T/1\s}}} \rp^2 
\right \}^{1/2}.
\eeq
}

For measurement times $T$ exceeding $\sim 0.1\s$, additional sources of noise not present in the
model discussed so far become dominant, obscuring 
the phase-diffusion noise. In order to distinguish between magnetic
and instrumental noise, we have measured fluctuations of ambient magnetic fields
with a fluxgate magnetometer external to the shields
at approximately $20\uG/\sqrt{\hz}$ 
above a $1/f$ corner frequency of approximately $0.3\hz$.
With a 
measured shielding factor of $3\times 10^5$, this implies a 
negligibly small white magnetic-field noise of around
$70\pG/\sqrt{\hz}$ at the vapor cell. 
The expected average noise on the supplied
bias field inside the shield 
is better than $6\nG/\sqrt{\hz}$ between $0.1\hz$ and $100\hz$.
This value is of the same order of magnitude as the observed Allan
deviation floor, although direct and reliable measurements of this
bias current have not been achieved. Noise on the bias-field current
could be distinguished from other sources by use of a magnetic 
gradiometer (see, for example, a description of a gradiometer based on a pair 
of FM NMOR sensors in Ref. \cite{xu2006}), though the small size of our magnetic shield has so
far precluded such a measurement.

Other  noise sources include
sensitivities of
the oscillation frequency to laser powers and detuning. A sensitivity to
pump or probe power arises, for instance, when the feedback-network 
phase shift deviates from the value which produces maximal
oscillation amplitude. Since the NMOR phase shift depends on the resonance
line width as in Eq.\ \eqref{phase_freq_rel}, while the line width depends
on optical power through the effect of power-broadening, changes in pump
or probe power produce 
changes in phase shift and corresponding 
deviations of the self-oscillation frequency. This effect vanishes to first order
precisely on the NMOR resonance; additional mechanisms for translating power
and detuning fluctuations into oscillation-frequency fluctuations are currently
being investigated.

\section{Conclusion}
We have demonstrated a self-oscillating two-beam magnetometer based 
on nonlinear magneto-optical rotation and shown that the independent
adjustment of pump and probe polarizations provides a powerful and frequency-independent
means of supplying the phase shift necessary for self-oscillation. Moreover,
the use of an unmodulated probe eliminates the necessity of elaborate filtering
procedures, producing instead a clean sine wave suitable for feeding back
as the self-modulation signal.
The resulting device possesses a high bandwidth and a measured sensitivity
of $3\nG$ at $1\s$. Considerable improvement, approaching the fundamental
atomic and optical shot-noise
limit of  $\apprle 30\pG$ in $1\s$ measurement time for a $\sim 6\cm^3$ cell, 
is expected through a more
thorough control of light-power-dependent, laser-frequency-dependent, and
electronic phase shifts, as well as through re-optimization 
using a quieter amplifier.
The fundamental constituents of this magnetometer are small and lightweight,
lending themselves well to designs for the field and for space. 
Operation in the geomagnetic range of fields has been achieved.
In future work,
the robustness of the magnetometer in an unshielded environment and an arbitrarily
directed geomagnetic field will be investigated. Performance in the presence
of splitting of the resonance line by the quadratic Zeeman shift will also be 
evaluated.

{
The authors acknowledge discussions with E. B. Alexandrov, S. M. Rochester 
and J. Kitching
and contributions of V. V. Yashchuk and J. E. Stalnaker to the design and 
construction of the magnetic shields, coils, and the vapor-cell heating system. 
This work is supported by DOD MURI grant \# N-00014-05-1-0406, 
by an ONR STTR grant through Southwest Sciences, Inc., and by an 
SSL Technology Development Grant.
}

\bibliographystyle{prsty}
\bibliography{kitching,budker,romalis,magnetometry,general,weis}

\end{document}